# Semi-Dirac dispersion relation in photonic crystals

Ying Wu

*Division of Computer, Electrical and Mathematical Sciences and Engineering, King Abdullah University of Science and Technology (KAUST), Thuwal 23955-6900, Saudi Arabia*

**Abstract**

A semi-Dirac cone refers to a peculiar type of dispersion relation that is linear along the symmetry line but quadratic in the perpendicular direction. Here, I demonstrate that a photonic crystal consisting of a square array of elliptical dielectric cylinders is able to produce this particular dispersion relation in the Brillouin zone center. A perturbation method is used to evaluate the linear slope and to affirm that the dispersion relation is a semi-Dirac type. Effective medium parameters calculated from a boundary effective medium theory not only explain the unexpected topological transition in the iso-frequency surfaces occurring at the semi-Dirac point, they also offer a perspective on the property at that point, where the photonic crystal behaves as a zero-refractive-index material along the symmetry axis but functions like at a photonic band edge in the perpendicular direction. Wave control, including beam splitting and directional beam shaping, is also demonstrated.

Email: ying.wu@kaust.edu.sa

**Introduction**

Photonic crystals (PhCs), periodic structures that control photons in a way comparable to the way semiconductors control electrons, have inspired extensive study since their emergence in the late 1980's [1,2]. Details about these fascinating structures can be found in a recent monograph [3], and references therein. One key feature of a PhC is the photonic band gap, which prevents light from propagating in certain directions at specific frequencies. It is analogous to the electron band gap in semiconductors. Engineering the photonic band gap to create a large and complete band gap was the goal of much early research in this area [1,2,4] . On the other hand, engineering the photonic bang gap to achieve a type of gapless band structure, namely the Dirac and Dirac-like cone dispersion relation, has been the focus of much recent work [5-23]. This dispersion relation is analogues to the Dirac cones in electron systems, where two linear bands touch so that there is no band gap. Owing to this special property, remarkable wave transport behaviors and interesting applications in electromagnetic waves have been reported [5-16]. To name a few, Haldane and Raghu proposed the possible realization of a directional optical wave guide in a PhC with breaking time-reversal symmetry [5] and described classical analogs of edge states in Quantum-Hall-Effect systems [6]. Sepkhanov *et al* reported the extremal transmission at the Dirac point [7]. Zhang *et al* observed the classical analogs of *Zitterbewegung* in photonic/phononic crystals with Dirac cones in the corner of the Brillouin zone [8]. Huang *et al* designed and fabricated a PhC-based zero-refractive-index material that possesses a Dirac-like cone in the Brillouin zone center and demonstrated both theoretically and experimentally the PhC's fascinating wave manipulation characteristics, such as beam shaping and cloaking of an object [9].

Various theoretical approaches, such as multiple scattering [18], tight binding [19], and perturbation [20-22], have also been developed to analyze the properties of Dirac cones in PhCs.

Very recently, a unique and unprecedented band structure was discovered in a $VO_2$/$TiO_2$ structure: near a point Fermi surface in the two-dimensional Brillouin zone, the dispersion relation is linear along the symmetry line ((1,1) direction) but quadratic in the perpendicular direction [23]. The associated quasiparticles are therefore massless along one direction, like those in graphene, but effective-mass-like along the other. This dispersion relation is called a semi-Dirac cone and the associated point is called a semi-Dirac point [23]. It was reported that such a point is associated with the topological phase transition between a semi-metallic phase and a band insulator [24]. Because both band gaps and Dirac cones in electron systems have found their classical analogies in PhCs, a natural question to ask is: if the band structure with the special peculiarity of a semi-Dirac cone can be transcribed into PhCs? If the answer to this question is yes, it is possible to envisage various interesting consequences that could be attributed to this unique dispersion relation, such as super anisotropic wave transport behavior near the semi-Dirac point.

Here, I demonstrate that by employing accidental degeneracy in a two-dimensional PhC with anisotropic scatterers, it is indeed possible to achieve a semi-Dirac type dispersion relation in the Brillouin zone center, which is associated with a semi-Dirac point. A $\vec{k} \cdot \vec{p}$ method [21, 22], based on perturbation, is exploited to confirm that the dispersion relation is linear in one symmetry axis (the $\Gamma X$ direction) and quadratic in the perpendicular one (the $\Gamma Y$ direction). Beyond that, the method shows that the linear slope decreases as $\vec{k}$ rotates away from the $\Gamma X$ direction and eventually vanishes when $\vec{k}$ is in the $\Gamma Y$ direction. A somewhat unexpected discovery is that the semi-Dirac point results in a topological transition [25] between a hyperbolic-like shape and an elliptic-like shape in the iso-frequency surfaces. An effective medium theory based on boundary integration uncovers the secret behind this topological transition, which is a result of different combinations of the signs in the effective medium parameters of an anisotropic medium. The theory also elucidates the key feature of the semi-Dirac point: it has a zero-refractive-index along the $\Gamma X$ direction while it is photonic-band-edge-like along the $\Gamma Y$ direction. Two wave manipulation examples, beam splitting and directional beam shaping, are demonstrated.

**The Photonic Crystal System and the Semi-Dirac Dispersion Relation**

The PhC considered in this study is a square array of elliptical cylinders with a dielectric constant of $\varepsilon_s = 12.5$ embedded in air ($\varepsilon_0 = 1$). The inset of Fig. 1(a) illustrates the unit cell of this PhC. The semi-minor axis of each elliptical cylinder is $r_a = 0.188a$, where $a$ is the lattice constant, and the semi-major axis is 1.3 times that of the semi-minor axis, i.e., $r_b = 1.3 r_a$. The electromagnetic wave is transverse electric (TE) polarized and its electric field, $\vec{E} = (0, 0, E_z)$, is always perpendicular to the plane of periodicity. Figure 1(a) shows the band structure of this PhC calculated by using COMSOL Multiphysics, a commercial package based on finite-element simulations. There exists a doubly degenerated point, marked as "A", in the Brillouin zone center at the dimensionless frequency $\tilde{\omega} = \omega a/2\pi c = 0.540$, where $\omega$ is the angular frequency and $c$ is the wave speed in air. This degenerated point is created by accidental degeneracy [9, 19-22] of a monopolar state (see Fig. 2 (b)) and a dipolar state (see Fig. 2(c)) at the Γ point, when the frequencies of these two states are deliberately tuned to be identical by adjusting the size or the material of the inclusion. Figures 1(b) and 1(c) show the band structure near the Γ point for smaller and larger dielectric cylinders, whose eccentricity is kept at 1.3 but their semi-minor axes become $0.180a$ and $0.195a$, respectively. Clearly seen in both figures are the separated modes at the Γ point marked as "A1" and "A2", where A1 indicates a dipolar mode and A2 corresponds to a monopolar mode. The reversed relative positions of points A1 and A2 for smaller and larger cylinders imply that there must be a case where A1 and A2 coincide, which, indeed, is the case when $r_a = 0.188a$ as discussed earlier. This accidental degeneracy produces a very interesting dispersion relation in the vicinity of this degenerated point. Roughly seen from Fig. 1(a) are two linear bands, along the ΓX direction, touching at Point A, and a quadratic band, along the ΓY direction, tangent to a flat band also at Point A. Below the flat band there is a directional gap in the ΓY direction.

**(I) Confirmation of the Semi-Dirac Dispersion: the $\vec{k} \cdot \vec{p}$ Method**

The unusual dispersion relation leads to several interesting questions: does the dispersion relation really behave like a semi-Dirac cone? Are the dispersion relations along the ΓX direction truly linear? If yes, can we predict their slopes? To answer these questions, I adopted a $\vec{k} \cdot \vec{p}$ method [21,22], based on perturbation theory, to study the dispersion relations, i.e., $\Delta\omega$ as a function of $\Delta k$, near Point A. The spirit of the $\vec{k} \cdot \vec{p}$ method is to take a set of eigenfunctions at a particular symmetry point ($\vec{k}_0$) of interest as a basis to study the eigenstates in the vicinity of that point. Accurate prediction of the linear slope of the dispersion relation is possible because the eigenfunctions takes all the multiple-scattering in the PhC into account. Here, the purpose is to discover the behavior of the dispersion relation near Point A. In principle, we should incorporate *all* the eigenstates at the Γ point ($\vec{k}_0 = 0$) as the basis. However, I considered only the doubly-degenerated states at Point A and a non-degenerated eigenstate marked as "B" in Fig. 1(a). The other bands are far away from Point A and only contribute to the higher order in $\Delta k$ [26]. Therefore, the task is to solve the following 3×3 secular equation:

$$\det \begin{vmatrix} \frac{\omega_{\vec{k}}^2 - \omega_B^2}{c^2} + P_{11} & P_{12} & P_{13} \\ P_{21} & \frac{\omega_{\vec{k}}^2 - \omega_A^2}{c^2} + P_{22} & P_{23} \\ P_{31} & P_{32} & \frac{\omega_{\vec{k}}^2 - \omega_A^2}{c^2} + P_{33} \end{vmatrix} = 0, \quad (1)$$

where $\omega_A$ ($\omega_B$) is the frequency of the eigenstates at Point A (B), and $P_{ij}$ represents the mode-coupling integrals among the three eigenstates whose profiles are plotted in Figs. 2(a)-2(c). Figure 2(a) shows the electric field distribution of the state located at Point B, where a dipolar state with its magnetic field parallel to the horizontal axis is seen. This state is denoted as $\psi_1$. Figures 2(b) and 2(c) exhibit the electric field map of the doubly degenerated states at Point A. Obviously seen are a monopolar state (Fig.2(b)) labeled as $\psi_2$ and a dipolar state with its magnetic field parallel to the vertical axis (Fig. 2(c)) labeled as $\psi_3$. By substituting $\psi_i$ ($i = 1,2,3$) into Eqs. (5)-(7) of Ref. [21], we can obtain the values of $P_{ij}$ to solve Eq. (1). Because I am interested in the region close to Point A,

I can express $\omega_{\vec{k}}$ as $\omega_{\vec{k}} = \omega_A + \Delta\omega_{\vec{k}}$. Thus, $\omega_{\vec{k}}^2 - \omega_A^2$ is approximated by $2\omega_A\Delta\omega_{\vec{k}}$ and $\omega_{\vec{k}}^2 - \omega_B^2$ becomes $\omega_A^2 - \omega_B^2$, which is a number that allows the 3×3 matrix to be downfolded to a 2×2 one. Given that the dimensionless frequency at Point A is $\tilde{\omega}_A = 0.540$, the solution to Eq. (1) is easily derived, which reads:

$$\Delta\tilde{\omega}_{\vec{k}} = (\pm 0.0459 \cos\beta)\Delta k + O(\Delta k^2), \tag{2}$$

where $\beta$ denotes the angle between $\vec{k}$ and the $\Gamma X$ direction, and the dimensionless frequency, $\Delta\tilde{\omega} = \Delta\omega a/2\pi c$, is used. This solution clearly points to an important conclusion that the dispersion relation near Point A has non-zero linear dependence in $\Delta k$ in *all* directions except for the $\Gamma Y$ direction where $\cos\beta$ vanishes. The linear slope, $\Delta\tilde{\omega}_{\vec{k}}/\Delta k$, takes the largest value when $\vec{k}$ is along the $\Gamma X$ symmetry axis, decreases as $\vec{k}$ rotates away from the $\Gamma X$ direction, and vanishes when $\vec{k}$ is along the $\Gamma Y$ symmetry axis. To validate Eq. (2), I present in Figs. 2(d) and 2(e) close-up views of the dispersion relations near Point A. The dots are the result of numerical simulations and the red lines are obtained by using Eq. (2). Good agreement between them is observed for the linear dispersions along the $\Gamma X$ and $\Gamma M$ directions. The dots in Fig. 2(d) clearly indicate a quadratic dispersion relation along the $\Gamma Y$ direction, which is predicted by Eq. (2) as well. The blue dashed curves in Fig. 2(d) are obtained by solving Eq. (1) to the second order in $\Delta k$. The upper blue curve fits the dots well when $\Delta k$ is small, but it gradually deviates from the dots as $\Delta k$ increases. The deviation is due to the fact that to get an accurate coefficient of $\Delta k^2$, we need to consider both the contributions of the far-away states [26] and the second-order perturbations, which are beyond the scope of the $\vec{k}\cdot\vec{p}$ method used here and are therefore ignored.

Straightforward linear algebra derivation reveals that the origin of the linear-parabolic dispersion relation resides in the strength of the mode-coupling integral between $\psi_2$ and $\psi_3$, i.e., $\int_{unit\ cell} \psi_2^*(\vec{r})\nabla\psi_3(\vec{r})d\vec{r}$. The linear term disappears as the integral vanishes. Along the $\Gamma Y$ direction, this integration is zero because the coupling between $\psi_2$ and $\psi_3$ is extremely weak. Even though the accidental degeneracy is achieved, no linear dispersion is therefore found [22]. In fact, if we examine the eigenstates, it is not difficult to find that the underlying physics lies in $\psi_3$, the dipolar mode of the doubly degenerated state. As shown in Fig. 2(c), the magnetic field of this dipolar mode is polarized vertically, implying it is a longitudinal mode along the $\Gamma Y$ direction but a transverse mode along the $\Gamma X$ direction. In electromagnetic waves, the longitudinal branch is localized and almost does not couple to the incident wave and other branches [9, 27]. Thus, a flat band associated with the longitudinal dipolar mode is found in the $\Gamma Y$ direction. However, since $\psi_3$ is transverse to the $\Gamma X$ direction, it easily couples to the incident wave and other branches. The coupling between $\psi_2$ and $\psi_3$ is therefore strong in the $\Gamma X$ direction and the linear dispersion relation is found.

**(II) The Topological Transition: Anisotropic Effective Medium Theory**

This interesting anisotropic dispersion is comparable to a semi-Dirac cone in the sense that dispersion is linear along the symmetry line but quadratic along the perpendicular direction. This property is seen again in Figs. 3(a) and 3(b), which present simulated three-dimensional dispersion surfaces, from different view angles, in the frequency regime from $\tilde{\omega}=0.45$ to 0.70. Only a half semi-Dirac cone is observed in the upper branch. The lower branch, though, preserves a similar linear-parabolic property to the upper one, is shaped like a roof, which is flat in one direction and bends down in the other directions. By considering the upper and lower surfaces together, I therefore call them a generalized semi-Dirac cone. Their kissing point is still called a semi-Dirac point. The iso-frequency surface contours for the upper and lower branches are plotted in Figs. 3(c) and 3(d), where hyperbolic-like and elliptic-like shapes are manifest, respectively. Surprisingly seen is that the topological transition [25] in the iso-frequency surface occurs at the semi-Dirac point.

Because the Bloch wave vector is very small near the Brillouin zone center, the unexpected topological transition at the semi-Dirac point can be understood from an effective medium perspective. Here I adopt a boundary effective medium approach that was originally developed for elastic waves in Ref. [28]. The effective medium parameters are computed by using the constitutive relations, i.e.,

$$\bar{D}_z = \varepsilon^{eff} \bar{E}_z, \text{ and } \begin{pmatrix} \bar{B}_x \\ \bar{B}_y \end{pmatrix} = \begin{pmatrix} \mu_x^{eff} & 0 \\ 0 & \mu_y^{eff} \end{pmatrix} \begin{pmatrix} \bar{H}_x \\ \bar{H}_y \end{pmatrix},$$

where the averaged fields are evaluated from the eigenstate fields on the boundaries of the unit cell. For example, for an eigenstate whose Bloch wave vector, $\vec{k}$, is along the $\Gamma X$ direction, the averaged electric and magnetic fields are respectively:

$$\bar{E}_z = \left( \int_0^a E_z(x=0) dy + \int_0^a E_z(x=a) dy \right) / 2a,$$

$$\bar{H}_y = \left( \int_0^a H_y(x=0) dy + \int_0^a H_y(x=a) dy \right) / 2a$$

and the averaged electric displacement and magnetic induction fields are respectively:

$$\bar{D}_z = \left(\int_0^a H_y(x=a)dy - \int_0^a H_y(x=0)dy\right)/(-i\omega a^2)$$

$$\bar{B}_y = \left(\int_0^a E_z(x=a)dy - \int_0^a E_z(x=0)dy\right)/(i\omega a^2).$$

Here the Maxwell equation, $\nabla \times \vec{B} = \dfrac{\partial \vec{D}}{\partial t}$ and $\nabla \times \vec{E} = -\dfrac{\partial \vec{B}}{\partial t}$, and the Stocks theorem are exploited. For the eigenstates with Bloch wave vectors in the ΓY direction, similar expressions can be obtained. Figure 4(a) shows the results of the effective medium parameters evaluated by this boundary integration method using the eigenstates highlighted by the solid dots shown in Fig. 1(a). Figure 4(a) demonstrates that the effective permittivity calculated from eigenstates along the ΓX and ΓY directions are identical, because the electric field is a scalar for TE polarized waves. However, the permeability is anisotropic owing to the vector nature of the in-plane magnetic field. From Fig. 4(a) we can conclude that $\mu_y^{eff}$ and $\varepsilon^{eff}$ cross zero simultaneously at the frequency of the semi-Dirac point, whereas $\mu_x^{eff}$ equals zero at a lower frequency $\tilde{\omega} = 0.487$. The topological transition in the iso-frequency surface of the PhC can indeed be attributed to these effective medium properties. In fact, the iso-frequency surface of an anisotropic material is described as:

$$\frac{k_x^2}{\mu_y} + \frac{k_y^2}{\mu_x} = \omega^2 \varepsilon. \tag{3}$$

The signs of the material parameters determine the topology of the iso-frequency surface [25]. For the effective medium of the PhC considered here, in the frequency regime between 0.487 and 0.540, $\mu_x^{eff} > 0$, $\mu_y^{eff} < 0$ and $\varepsilon^{eff} < 0$. Such a combination of signs in the effective medium parameters not only leads to hyperbolic iso-frequency surfaces, which are close to those in Fig. 3(c), but it also gives rise to a band gap along the $\Gamma Y$ direction and a negative band along the $\Gamma X$ direction. This is because $k_y$ is purely imaginary when $k_x = 0$ while $n_x^{eff} = \sqrt{\varepsilon^{eff} \mu_y^{eff}} < 0$ when $k_y = 0$. Similar analysis can be applied to the higher branch when the frequency is above the semi-Dirac point, where all the material parameters are positive. Elliptic-like iso-frequency surfaces are therefore expected. The simultaneous zero $\mu_y^{eff}$ and $\varepsilon^{eff}$ achieved by accidental degeneracy leads to a linear dispersion relation along the $\Gamma X$ direction in the vicinity of the semi-Dirac point, whereas a single zero in $\varepsilon^{eff}$ and a positive $\mu_x^{eff}$ makes the dispersion relation quadratic along the $\Gamma Y$ direction. All the behaviors predicted by the effective medium parameters are in line with the properties of the simulated band structures.

**(III) Properties of the Semi-Dirac Dispersion Relation and Applications**

The effective medium theory reveals an important feature of the semi-Dirac-like point studied here. At this point, the PhC exhibits both a double-zero-refractive-index (permittivity and permeability simultaneously zero) material property and photonic band edge characteristics. Figures 4(b)-4(d) show the electric field pattern of a plane wave with the frequency of a semi-Dirac point impinging on a PhC slab inside a waveguide. The waveguide has boundary walls that are perfect magnetic conductors (PMC). Obviously, when the incident wave is propagating in the $\Gamma X$ direction, the PhC exhibits the typical transmission property of a double-zero-refractive-index material [9], i.e., total transmission is supported without any phase change inside the material as shown in Figs. 4(c) and 4(d) for the real and imaginary parts of the electric field, respectively. However, when the PhC is illuminated by the same incident wave along the Γ$Y$ direction, the amplitude of the electric field decays as the wave penetrates into the PhC, which is similar to the transmission property at the edge of a photonic band gap. This anisotropic transport feature provides evidence that the PhC has both the property of a "zero-index-material" and a "photonic-band-gap" material at the semi-Dirac point.

The anisotropic transport property and the topological transition in the iso-frequency surfaces lead to interesting wave manipulation behaviors. In Fig. 5, I demonstrate the radiation properties of a square sample with 16 by 16 rods that are illuminated by a point source located at the center of the sample at two different frequencies. Figure 5(a) and 5(b) show, respectively, the electric field and the flux distributions when the point source radiates at a dimensionless frequency 0.520. Due to the hyperbolic-like dispersion relation at that frequency, the out-going wave splits into four beams, that are consistent with the iso-frequency surface. However, when the frequency is slightly above the semi-Dirac point, i.e., $\tilde{\omega}=0.544$, the outgoing beams mainly go along the horizontal direction and the wave front is almost parallel to the vertical surfaces as shown in Fig. 5(c). The field pattern from the same source but inside a homogenous anisotropic medium is plotted in Fig. 5(d). Almost the same field pattern as that in Fig. 5(c) is observed. Here the effective medium parameters of the homogeneous medium are obtained from Fig. 4(a). They are $\varepsilon^{eff}=0.018$, $\mu_y^{eff}=0.042$, and $\mu_x^{eff}=0.495$.

**Conclusions**

In conclusion, I have demonstrated, by accidental degeneracy, that a two-dimensional PhC comprising a square array of elliptical dielectric cylinders exhibits a semi-Dirac dispersion relation and a semi-Dirac point in the Brillouin zone center. The $\vec{k} \cdot \vec{p}$ method affirms that in the vicinity of the semi-Dirac point the dispersion relation is linear along one symmetry axis (the $\Gamma X$ direction) and quadratic along the perpendicular one (the $\Gamma Y$ direction). The semi-Dirac point is associated with a topological transition in its iso-frequency surfaces, which is explained by effective medium theory for an anisotropic medium. The semi-Dirac point is shown to behave as a zero-refractive-index material along the $\Gamma X$ direction and a photonic band edge material along the $\Gamma Y$ direction. Interesting wave manipulation properties, such as beam splitting and directional beam shaping, have been demonstrated.

**Acknowledgements**

The author would like to thank Prof. Z.Q. Zhang, Prof. C. T. Chan, Prof. J. Li, Prof. J. Mei and Dr. X. Q. Huang for fruitful discussions. Special thanks go to Prof. P. Sheng, Prof. Y. Lai, Prof. Z. H. Hang and Prof. M. H. Lu for their comments. I acknowledge V. Unkefer for editorial work on this manuscript. This work was supported by KAUST's Baseline Research Fund.

**References**

[1] Eli Yablonovitch, Phys. Rev. Lett. **58**, 2059 (1987)

[2] S. John, Phys. Rev. Lett. **58**, 2486 (1987).

[3] J. D. Joannopoulos, S. Johnson, J. Minn, and R. Meade, *Photonic Crystals: Molding the Flow of Light* 2$^{nd}$ Ed. (Princeton University Press, Princeton and Oxford, 2008)

[4] K. M. Ho, C. T. Chan, and C. M. Soukoulis, Phys. Rev. Lett. **65**, 3152 (1990)

[5] F. D. M. Haldane and S. Raghu, Phys. Rev. Lett. **100**, 013904 (2008);

[6] S. Raghu and F. D. M. Haldane, Phys. Rev. A **78**, 033834 (2008).

[7] R. A. Sepkhanov, Y. B. Bazaliy, and C. W. J. Beenakker, Phys. Rev. A **75**, 063813 (2007); R. A. Sepkhanov and C. W. J. Beenakker, Opt. Commun. **281**, 5267 (2008).

[8] X. Zhang, Phys. Rev. Lett. **100**, 113903 (2008); X. Zhang and Z. Liu, *ibid* **101**, 264303 (2008); X. Zhang, Phys. Lett. A **372**, 3512 (2008).

[9] X. Huang, Y. Lai, Z. H. Hang, H. Zheng, and C. T. Chan, Nat. Mater. **10**, 582 (2011)

[10] R. A. Sepkhanov, J. Nilsson, and C. W. J. Beenakker, Phys. Rev. B **78**, 045122 (2008).

[11] T. Ochiai and M. Onoda, Phys. Rev. B **80**, 155103 (2009)

[12] M. Diema, T. Koschny, and C. M. Soukoulis, Physica B, **405**, 2990 (2010)

[13] V. Yannopapas, Phys. Rev. B **83** 113101 (2011)

[14] J. Bravo-Abada, J. D. Joannopoulos, and M. Soljačić, Proc. Natl. Acad. Sci. **109**, 9761 (2012)

[15] A. B. Khanikaev, S. H. Mousavi, W-K Tse, M. Kargarian, A. H. MacDonald, and G. Shvets, Nat. Mater. **12**, 233 (2013)

[16] M. C. Rechtsman, J. M. Zeuner, Y. Plotnik, Y. Lumer, D. Podolsky, F. Dreisow, S. Nolte, M. Segev, and A. Szameit, Nature, **496**, 196 (2013)

[17] Y. P. Bliokh, V. Freilikher, and F. Nori, Phys. Rev. B **87** 245134 (2013)

[18] C. T. Chan, Z. H, Hang, and X. Q. Huang, Advances in OptoElectronics, 2012 313984 (2012)

[19] K. Sakoda, Opt. Express, **20**, 3898 (2012);

[20] K. Sakoda, Opt. Express, **20**, 25181 (2012);

[21] J. Mei, Y. Wu, C. T. Chan, and Z. Q. Zhang, Phys. Rev. B **86**, 035141 (2012)

[22] Y. Li, Y. Wu, X. Chen, and J. Mei, Opt. Express **21**, 7699 (2013)

[23] V. Pardo and W. E. Pickett, Phys. Rev. Lett. **102**, 166803 (2009); S. Banerjee, R. R. P. Singh, V. Pardo,and W. E. Pickett, *ibid* **103**, 016402 (2009)

[24] M. O. Goerbig, Rev. Mod. Phys. **83**, 1193 (2011) G. Montambaux, F. Piéchon, J.-N. Fuchs, and M. O. Goerbig, Phys. Rev. B **80**, 153412 (2009)

[25] H. N. S. Krishnamoorthy, Z. Jacob, E. Narimanov, I. Kretzschmar, and V. M. Menon, Science **336**, 205 (2012)

[26] B. A. Foreman, J. Phys.: Condens. Matter **12**, R435 (2000).

[27] Y. Wu, J. Li, C. T. Chan, and Z. Q. Zhang, Phys. Rev. B **74**, 085111(2006)

[28] Y. Lai, Y. Wu, P. Sheng, and Z. Q. Zhang, Nat. Mater. **10**, 620 (2011)

**Figures**

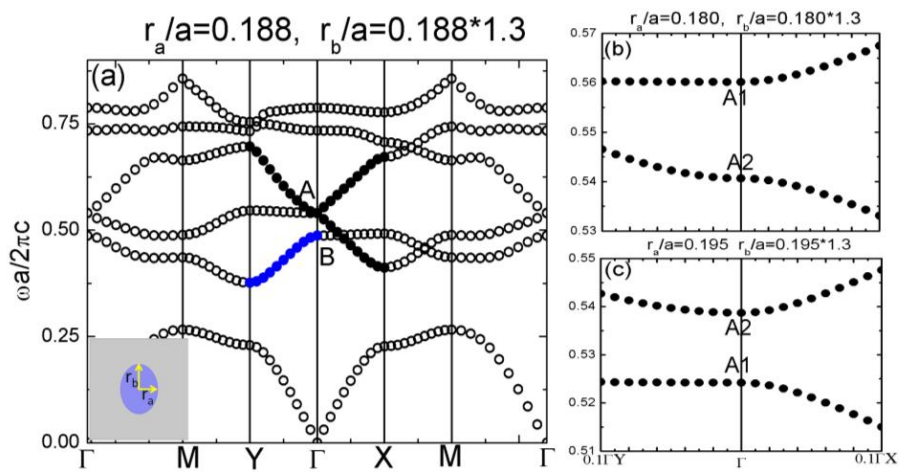

**Figure 1** (a) The band structure of a two-dimensional photonic crystal comprising a square array of elliptical cylinders with a relative dielectric constant, $\varepsilon_s = 12.5$, embedded in air. The lattice constant is $a$, which also serves as the length unit. The semi-minor axis, $r_a$, is $0.188a$, and the semi-major axis is 1.3 times that of the semi-minor axis. A doubly-degenerated state in the Brillouin zone center is found near the dimensionless frequency, 0.540, marked as "A". In the vicinity of this point, the dispersion relation is linear along the $\Gamma X$ direction and quadratic along the $\Gamma Y$ direction, which is shown more clearly in Fig. 2(d). (b) and (c) Enlarged views of the band structure for smaller and larger elliptical cylinders. The doubly-degenerated state shown in (a) splits into two single states, marked as A1 and A2, where A1 corresponds to a dipolar state and A2 corresponds to monopolar state.

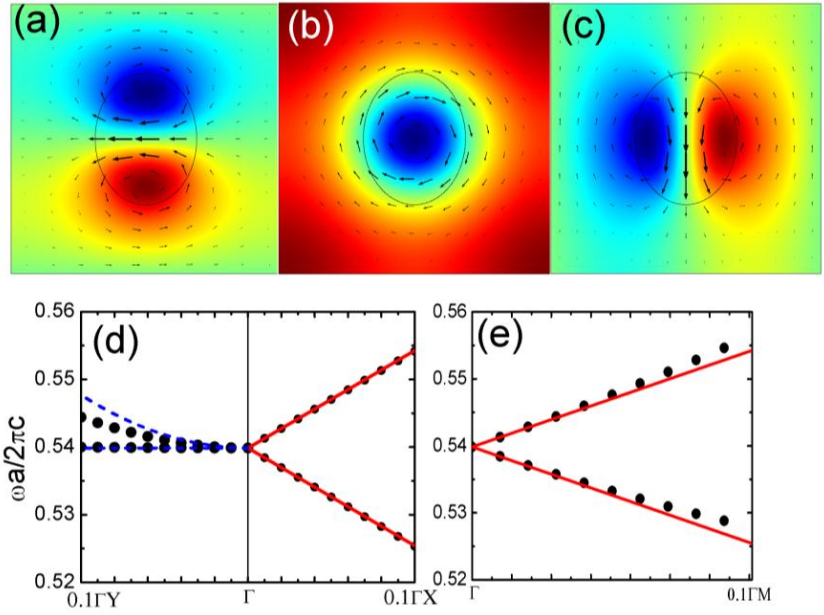

**Figure 2** (a) The electric field pattern of the eigenstate marked as "B" in Fig. 1(a). Dark red and dark blue indicate the maximum positive and negative values, respectively. This is a dipolar state with a magnetic field parallel to the x axis, indicated by the arrows. (b) and (c) The electric field patterns of the doubly-degenerated state marked as "A" in Fig. 1(a). A monopolar and a dipolar state with the magnetic field (arrows) perpendicular to the x axis are seen. (d) An enlarged view of the band structure near the doubly-degenerated state. The dots are calculated by COMSOL. Linear dispersion is seen along the $\Gamma X$ direction, while a quadratic dispersion relation is manifest along the $\Gamma Y$ direction. Red solid lines and blue dashed curves are obtained from the $\bar{k} \cdot \bar{p}$ method, which verifies the linear dispersion relation. (e) The same as (d) but along the $\Gamma M$ direction. The linear dispersion relation is seen again.

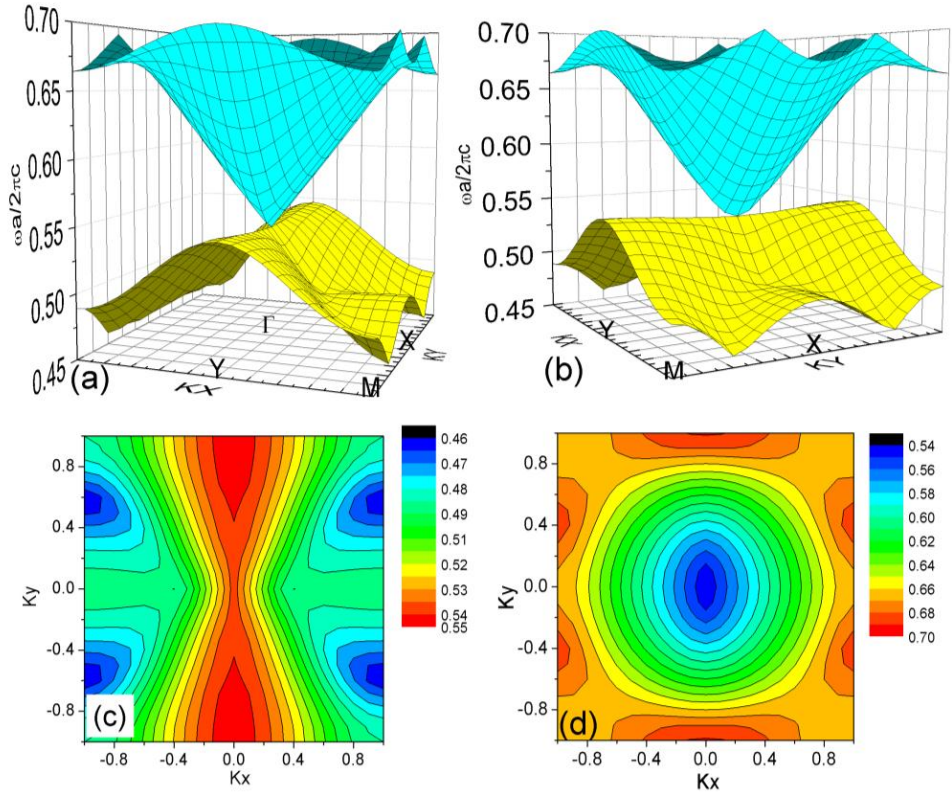

**Figure 3** (a) and (b) The three-dimensional band structure of the photonic crystal. The upper surface is a semi-Dirac cone. Near its bottom, it is linear in frequency along all directions except for the $\Gamma Y$ direction, which is quadratic. It touches the lower surface at the Brillouin zone center near the dimensionless frequency 0.54. The lower surface is flat in one direction and bends down along the other directions. (c) and (d) The iso-frequency surfaces of the lower and higher branches, where a hyperbolic-like and elliptic-like surfaces are found, respectively.

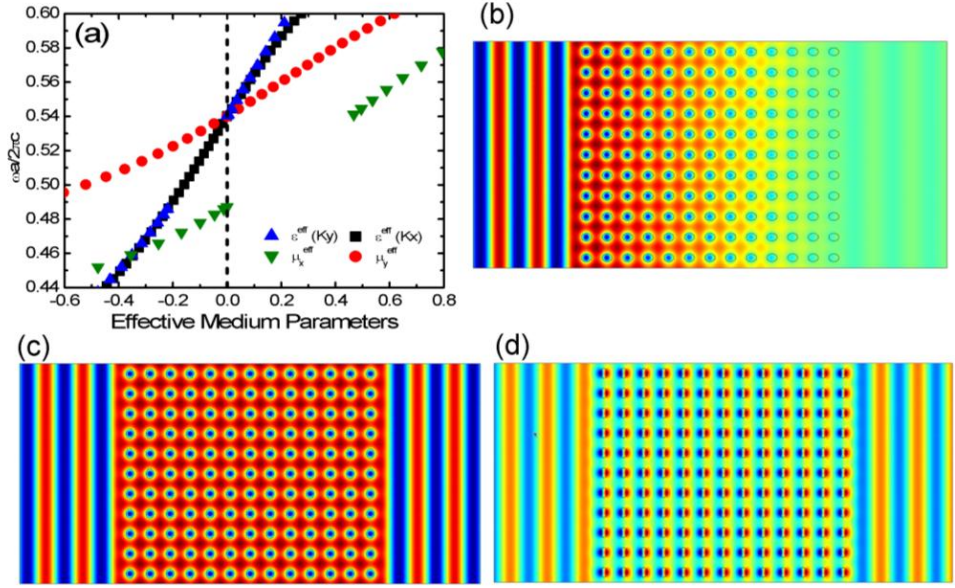

**Figure 4** (a) Effective medium parameters evaluated with a boundary effective medium theory using the eigenstates highlighted by solid dots in Fig. 1(a). Blue upper triangles and black squares represent the effective permittivity calculated by using eigenstates along the $\Gamma Y$ and $\Gamma X$ directions, respectively. They almost overlap indicating that the effective permittivity, $\varepsilon^{eff}$, is a scalar and does not depend on the direction. Red circles are $\mu_y^{eff}$, which cross zero at the same point as $\varepsilon^{eff}$ and the point is a semi-Dirac point. Green lower triangles represent $\mu_x^{eff}$, which cross zero at dimensionless frequency 0.487. Note that both blue upper triangles and green lower triangles are missing from the frequency regime at 0.487 to 0.540, which corresponds to a band gap along the $\Gamma Y$ direction. No eigenstates are thus available to evaluate the related effective medium parameters. (b)-(d) The electric field for a plane wave impinging on a PhC slab in a waveguide whose walls have PMC boundary conditions at the semi-Dirac frequency 0.540. (b) The real part of the electric field when the incident wave is along $\Gamma Y$ direction. The amplitude decays as the wave penetrates into the sample. The imaginary part is orders of magnitude smaller than the real part, which is why it is not shown here. (c) and (d) The real and imaginary parts of the electric field when the incident wave is along the $\Gamma X$ direction. Both suggest that there is no phase change in the sample, which is a typical property of a double-zero-index material.

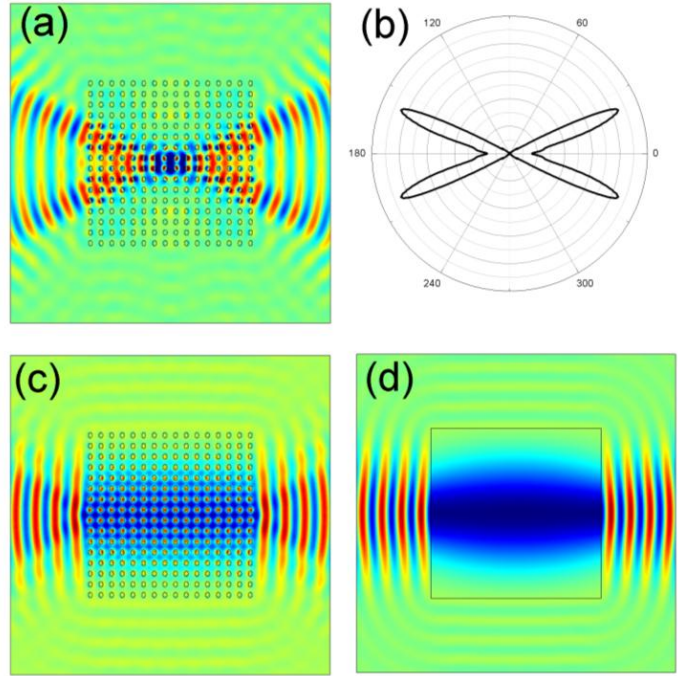

**Figure 5** A point source is placed inside the center of a square sample of 16 by 16 rods. (a) and (c) show the electric field patterns when the source frequency is below (0.520) and slightly above (0.544) the semi-Dirac point, respectively. Beam splitting and directional beam shaping is observed. (b) The radial flux as a function of the angle for the case simulated in (a). (d) The same as (c) but the sample is replaced by its effective medium, whose parameters are $\varepsilon^{eff}=0.018$, $\mu_y^{eff}=0.042$, and $\mu_x^{eff}=0.495$. A similar pattern to that shown in (c) is found.